\documentclass[aps,prl,twocolumn,groupedaddress,showpacs]{revtex4}

\usepackage{graphicx}
\usepackage{dcolumn}
\usepackage{bm}
\usepackage{subfigure}
\usepackage{amsmath}
\usepackage{amssymb}

\begin{document}

\preprint{}

\title{Frequency evaluation of the doubly forbidden $^1S_0\rightarrow\,^3P_0$ transition in bosonic $^{174}$Yb}

\author{N. Poli}
\altaffiliation{LENS and Dipartimento di Fisica, Universit\`a di
Firenze, INFN - sezione di Firenze - 50019 Sesto
Fiorentino, Italy}
\author{Z. W. Barber}
\author{N. D. Lemke}
\altaffiliation{also at University of Colorado, Boulder, CO, 80309}
\author{C. W. Oates}
\author{L. S. Ma}
\altaffiliation{State Key Lab. of Precision Spectroscopy, ECNU, China}
\author{J. E. Stalnaker}
\altaffiliation{present address: Department of Physics and Astronomy, Oberlin College, Oberlin OH 44074}
\author{T. M. Fortier}
\author{S. A. Diddams}
\author{L. Hollberg}
\author{J. C. Bergquist}
\author{A. Brusch}
\author{S. Jefferts}
\author{T. Heavner}
\author{T. Parker}

\affiliation{National Institute of Standards and Technology\\325
Broadway, Boulder, CO 80305}
\thanks{Official contribution of the National Institute of Standards and Technology of the U.S. Department of Commerce; not subject to copyright.}
\date{\today}

\begin{abstract}
We report an uncertainty evaluation of an optical lattice clock based on the $^1S_0\leftrightarrow\!^3P_0$ transition in the bosonic isotope $^{174}$Yb by use of magnetically induced spectroscopy. The absolute frequency of the $^1S_0\leftrightarrow\!^3P_0$ transition has been determined through comparisons with optical and microwave standards at NIST. The weighted mean of the evaluations is $\nu$($^{174}$Yb)=518\,294\,025\,309\,217.8(0.9)~Hz. The uncertainty due to systematic effects has been reduced to less than 0.8~Hz, which represents $1.5\times10^{-15}$ in fractional frequency.  
\end{abstract}

\pacs{32.10.Dk, 06.30.Ft, 32.70.Jz, 39.30.+w}
\maketitle
Optical frequency references are now reaching fractional frequency uncertainties lower than those of the best Cs primary standards, which define the second.  Intense efforts to reduce these levels to well below one part in $10^{16}$ are underway with many different atomic systems (neutral and ionic) for a variety of applications including tests of fundamental physics and the development of the next generation of primary frequency standards~\cite{Rosenband08,Ludlow08}.  Here we report the first detailed evaluation and absolute frequency measurement (with an uncertainty of $1.5\times10^{-15}$) of a clock system based on neutral $^{174}$Yb atoms confined to a one-dimensional optical lattice.  Ytterbium is an attractive atom for optical clock studies due to its large mass and numerous abundant isotopes that offer a variety of nuclear spins (0,1/2,5/2).  For this work, we focus on the potential of spin-zero isotopes for lattice clock work with the first evaluation of such a system at the $1\times10^{-15}$ level.

Spin-zero isotopes are appealing for frequency metrology due to their simple structure.  Indeed atomic clocks based on $J=0\leftrightarrow J'=0$ transition in these isotopes present a nearly pure two-level system in that it lacks Zeeman structure, first-order sensitivity to magnetic fields, optical pumping effects, and first-order vector/tensor sensitivity to the lattice laser intensity and polarization.  Moreover, the spin-zero (even) isotopes in the alkaline earth-like atom systems are generally more abundant and easier to prepare (through laser cooling and trapping) for high resolution spectroscopy.  To realize these advantages, however, it is necessary to use an external field such as a magnetic bias field (as in this work) to induce a nonzero transition strength for the otherwise truly forbidden transition~\cite{Taichenachev06, Barber06}.  The presence of this external field leads to a small shift of the clock frequency that must be calibrated.  Additionally, the weakness of the induced transition requires higher probe light intensities than are used with the odd (half-integer spin) isotopes, which lead to an AC Stark shift that also requires careful evaluation.  This is the essential trade-off between the odd and even isotopes: one accepts two larger systematic effects in exchange for a considerably simpler atomic system that does not require optical pumping or hopping between multiple spectroscopic features in order to suppress first-order magnetic field and lattice polarization sensitivities.  Optical lattice clock studies with Sr have thus far been focused on the odd isotope ($^{87}$Sr, with nuclear spin 9/2), and the one evaluation with an even isotope ($^{88}$Sr) had an uncertainty considerably higher than those of its odd counterparts~\cite{Baillard07}.  Here we show that the even isotopes, especially in the case of Yb, which has somewhat more favorable atomic properties for even isotope work~\cite{Taichenachev06}, can achieve extremely low uncertainties as well, with no identified barriers to reaching the $10^{-17}$ level.  

Experimental details on the cooling and trapping setup have been reported in~\cite{Barber07}.  In brief, $^{174}$Yb atoms are trapped and cooled in a two-stage magneto-optical trap (MOT) working on the strong dipole allowed $^1S_0\leftrightarrow\!^1P_1$ transition at 398.9 nm and then on the intercombination $^1S_0\leftrightarrow\!^3P_1$ transition at 555.8~nm.  Typically 10$^5$ atoms are trapped at a temperature of less than $40~\mu$K in 350~ms. About $10^4$ atoms are then loaded to a 1D standing wave optical lattice operating at the magic wavelength of 759.35~nm for the $^1S_0\leftrightarrow\!^3P_0$ clock transition. A lattice potential depth of about 500~$E_r$ (recoil energy $E_r/k_B$ = 100~nK for Yb at the lattice frequency) is realized by tightly focusing (waist $\approx30$~$\mu$m) a 1~W injection-locked Ti:Sapphire laser.  After a 25~ms delay that allows for the quadrupole magnetic MOT field to dissipate and the static magnetic field to be turned on for the spectroscopy, the $^1S_0\leftrightarrow\!^3P_0$ clock transition at 578~nm is probed with a $\pi$-pulse lasting 50 to 150~ms.  The probe laser is collinear with the lattice beam and is focused onto the atoms with a waist of 60~$\mu$m.  Both the lattice polarization and the probe laser polarization are aligned with the vertical static magnetic field.

The 578~nm clock laser is based on sum-frequency generation of a Nd:YAG laser at 1.32~$\mu$m, and an Yb fiber laser at 1.03~$\mu$m~ in a periodically-poled lithium niobate waveguide~\cite{Oates07}. The 578~nm light produced is then frequency stabilized with a Pound-Drever-Hall lock to a vertically-mounted high finesse stable cavity. The short-term frequency stability is $2.2\times10^{-15}$ at 1 to 2 s, with a residual frequency drift of less than 0.4~Hz/s. The residual drift is monitored and canceled to first order through feed-forwarding to an acousto-optical modulator interposed between the laser source and the reference cavity. The stabilized yellow light is then sent through phase-noise-compensated fibers to both a femtosecond frequency comb for clock comparison and to the atom trap.  For a typical magnetic field strength of $B_0 = 1.3$~mT and a probe intensity of $I_0 = 300$~mW/cm$^2$, the Rabi frequency is $\Omega\sim5$~Hz, which generates a Fourier-limited linewidth of about $\Delta\nu = 10$~Hz. To prevent spurious ac Stark shifts, the pre-cooling and trapping beams are switched off during the spectroscopy with acousto-optical modulators and mechanical shutters.

The amount of population transfer via the clock transition is determined through fluorescence detection on the strong 399~nm transition of the atoms remaining in the $^1S_0$ ground state.  To lock the frequency of yellow probe laser to the atomic transition, an error signal is derived by alternately probing the half-width points and then demodulating with a digital microprocessor, which provides frequency corrections to an acousto-optical modulator.  For a typical observed S/N ratio of $\sim10$ and a probe duty cycle of 15~\%, the projected clock stability is $1\times10^{-15}\,\tau^{-1/2}$.

\begin{figure}
\includegraphics[height=6.5cm]{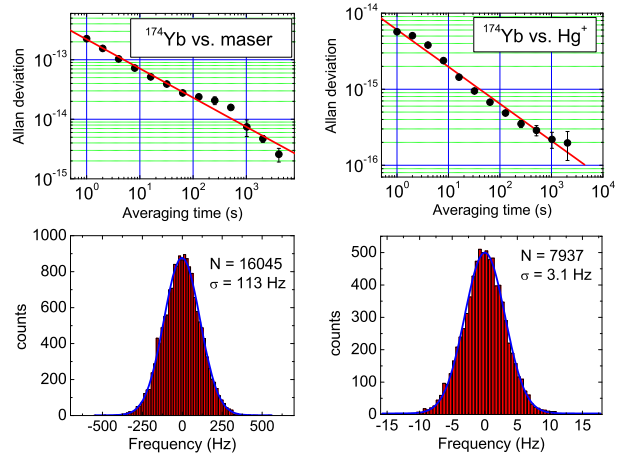}
\caption{Frequency comparison of $^{174}$Yb against a maser calibrated to NIST-F1 (left) and Hg$^+$ (right) through use of the optical fequency comb~\cite{Stalnaker07,Fortier06}.  Histograms and Allan deviation generated from series of 1~s gate interval frequency counts.  The linear fit on the two Allan deviations give a slope of $2\times$10$^{-13}~\tau^{-1/2}$ and $5.5\times$10$^{-15}~\tau^{-1/2}$ respectively.  The statistical uncertainties for the two measurements are 0.9~Hz and 0.10 Hz.}
\label{figmeas}
\end{figure}

Evaluation of the systematic shifts of the Yb lattice clock was accomplished through comparisons with other optical frequency standards at NIST.  While initial measurements were performed against a neutral Ca standard~\cite{Oates00}, the measurements reported here were performed against the Hg$^+$ optical standard~\cite{Stalnaker07}.  For the comparisons an octave-spanning Ti:Sapphire optical frequency comb~\cite{Fortier06} was locked to the Hg$^+$ standard, and the beat frequency between the comb and the Yb standard was counted with 1~s gate times.  We typically changed the value of the experimental parameter under study every 200 to 300 s in order to determine the eventual shift with a statistical uncertainty of about $3\times10^{-16}$ (see Fig. \ref{figmeas}).

\begin{figure}
\includegraphics[height=6.5cm]{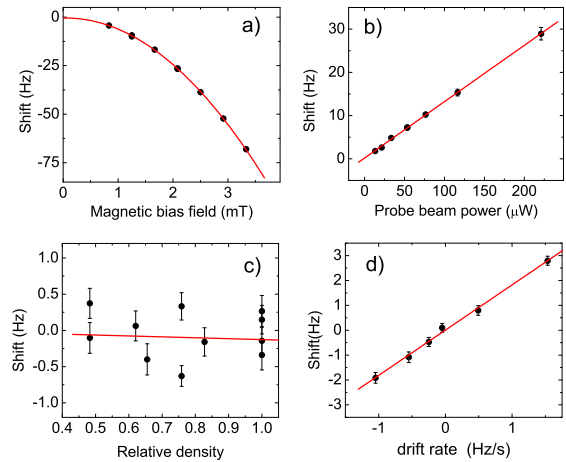}
\caption{Systematic shifts of the $^1S_0\leftrightarrow\!^3P_0$ transition in Yb optical lattice clock.  a) Measurement of the second order Zeeman shift. The second order coefficient is evaluated $\beta = -6.12(10)$~Hz/mT$^2$.  b) Probe ac linear Stark shift. The coefficient $\kappa = 15(3)$~mHz/(mW/cm$^2$) is in agreement with the theoretical estimation (see text).  c) Density shift.  The data points are consistent with zero density shift with a total deviation of 0.6~Hz. d) Servo error.  The frequency offset due to uncompensated linear drift of the reference cavity is shown (slope = 1.82(10)~Hz/(Hz/s)).  With periodic manual correction of the feed-forward system used to remove the linear cavity drift this uncertainty can be made negligible.}
\label{systmeas}
\end{figure}

In the case of spectroscopy of the clock transition with the even isotope, the linear Zeeman effect is zero, with only a second-order dependence on the magnetic field amplitude.  In Fig.~\ref{systmeas}a we show an evaluation of the second-order Zeeman shift obtained by varying the static magnetic field between 1~mT and 3.5~mT.  The resulting second-order Zeeman coefficient in $^{174}$Yb is $\beta = -6.12(10)$~Hz/mT$^2$, found by a quadratic fit to zero field.  For a field value of $1.6$~mT the shift is approximately 18~Hz and is determined with an uncertainty of less than $0.3$~Hz.  The uncertainty in the systematic shift ($\Delta_{\textbf{B}} = 2\beta|\textbf{B}|\delta\textbf{B}$) is limited mainly by the knowledge of the applied static magnetic field $B$, which is calibrated to better than 10~$\mu$T through spectroscopy of the magnetically sensitive $^1S_0\leftrightarrow\,^3P_1$ transition.  Strategies for reducing the uncertainty of the second-order Zeeman shift include refining the absolute calibration of the magnetic field down to less than 1~$\mu$T (one possibility could be the use of the narrower $^1S_0\leftrightarrow\!^3P_0$ transition in the fermionic $^{171}$Yb isotope) and operation at lower bias field values.

In a similar way, the ac Stark shift induced by the yellow probe light has been determined by measuring the shift as a function of intensity and extrapolating to zero intensity.  Although the probe laser power is controlled with good precision ($< 1$\%) with a photodiode and fast servo control, the absolute intensity at the location of the atoms is not well determined.  In the current setup, the clock excitation light is focused to a $\sim60$~$\mu$m waist at the atoms, which is larger by only about a factor of two than the waist size of the lattice beam.  This implies that the atoms experience an average intensity and not uniform illumination.  In addition, drifts in alignment can cause the average intensity experienced by the atoms to vary by 10~\% from day to day in the current setup.  Fortunately, the intensity experienced by the atoms is directly proportional to the optical power, so the shift due to the yellow probe can be accurately extrapolated to zero intensity.  As is reported in Fig.~\ref{systmeas}b, this effect shows a linear dependence on the probe intensity with a slope of $\kappa\approx15(3)$~mHz/(mW/cm$^2$), in agreement with the theoretical estimate~\cite{Taichenachev06}. In this case, the typical $\sim$300~mW/cm$^2$ probe intensity gives a total shift of about 7~Hz with an uncertainty of 0.2~Hz.  This uncertainty could best be lowered by illuminating the atoms with a larger beam and/or operating at lower values of the intensity.

It is important to emphasize that the present uncertainties of these two shifts do not pose fundamental limits for future evaluations.  Both the value of the shifts and the consequent errors scale linearly with linewidths.  In particular, an experimental linewidth of about 1~Hz could be produced with a reduced value of both the static magnetic field $B = 0.3$~mT and the probe light intensity $I = 70$~mW/cm$^2$. In this case, with the present level of field calibrations the fractional frequency uncertainty would be less than $1\times10^{-16}$, and a feasible improvement by a factor of 10 in the calibrations would reduce the overall uncertainty to less than $1\times10^{-17}$.

The frequency uncertainties due to lattice light have been studied in detail~\cite{Barber08}.  The current systematic uncertainties (polarizability and hyperpolarizability) due to the optical lattice (magic wavelength $\lambda_{magic}=759.3538$~nm) at a depth of 500\,$E_r$ are given in Table \ref{syst}, and considerable reduction of both is anticipated.

\begin{table}[h] \caption{Frequency uncertainty budget for the $^{174}$Yb optical lattice clock.  Some frequency shift values and their uncertainties depend on operating conditions and particular evaluation; typical values are given.  }
\label{syst}
\begin{tabular}{c c c}
\toprule Effect &  Shift (10$^{-15}$)& Uncertainty (10$^{-15}$)\\
\hline
2$^{nd}$ order Zeeman &  -18 -- -36 & 0.4\\
Probe light &  6 -- 12.0 & 0.4 \\
Lattice Polarizability & $<$ 1 & 0.6\\
Hyperpolarizability &  0.33 & 0.07\\
Density  &  -0.2 & 1.0 \\
Blackbody shift & -2.5 & 0.25 \\
\hline
Syst. Total &  -15 -- -27 &  $<$ 1.5\\
\botrule
\end{tabular}
\end{table}

An important systematic effect to be accounted for is any density shift due to cold collisions between atoms in the 1D optical lattice~\cite{Ludlow08}.  This has been evaluated by reducing the total atom number captured in the 399~nm MOT and subsequently the lattice, changing the density in the range 0.5-1$\times\rho_0$ where $\rho_0\sim10^{11}$~cm$^{-3}$ is the estimated mean density in the lattice (see Fig.~\ref{systmeas}c).  The result, -0.1(0.6)~Hz, is consistent with zero shift at this level of precision.  The conservative uncertainty quoted in Table \ref{syst} is due to the small range of densities for which the system has been run reliably and the current operation of the clock at the maximum atom number.  Further reduction in the uncertainty of the density shift will occur with better evaluations in a 1D lattice or through the use of an under-filled 2D or 3D lattice, for which any such shift should vanish. 

Finally, the Stark shift induced by blackbody radiation (BBR) has been estimated by measuring the mean value of the temperature of the MOT chamber (295(3)~K) and calculating the shift from the theoretical estimates in~\cite{Porsev06}.  The uncertainty is limited by knowledge of Yb polarizability at the 10~\% level, or $2.5\times10^{-16}$ fractionally.  Precise measurements of the BBR shift or operation at cryogenic temperatures will be required to achieve sub-$10^{-17}$ uncertainty.    

Table \ref{syst} summarizes the systematic frequency shifts for magnetic field induced spectroscopy on $^1S_0\leftrightarrow\!^3P_0$ transition in $^{174}$Yb.  The values of the individual shifts depend on the particular operating conditions chosen (i.e., lattice power and frequency, bias magnetic field strength, and probe laser intensity).  The quoted uncertainties of each frequency bias represent typical values.  The actual uncertainties for a given comparison depend on the operating conditions and the proximity to, and quality of, systematic shift calibrations. With improvements in the experimental setup and diagnostics, reduced long-term variations in operation conditions will allow for less frequent systematic shift calibrations.

Four absolute determinations of the clock transition frequency in $^{174}$Yb were performed over several months against a maser calibrated with the NIST-F1 primary frequency standard~\cite{Heavner05} or against the optical Hg$^+$ standard using the optical frequency comb~\cite{Stalnaker07,Fortier06}.  As shown in Fig. \ref{figmeas}, the noise contribution for both comparisons is mainly white in frequency, with a fractional instability of $2 \times 10^{-13}~\tau^{-1/2}$ vs. the maser and $5.5 \times 10^{-15}~\tau^{-1/2}$ vs. Hg$^+$.  Recent improvements to the 578\,nm stabilization cavity have resulted in an improvement of the Yb vs. Hg$^+$ fractional instability to less than $3\times10^{-15}~\tau^{-1/2}$, with the stability reaching to below $1\times10^{-16}$ in a couple of thousand seconds.  For the comparisons against the maser, statistical and calibration uncertainties are significant, and contribute a fractional uncertainty of about $2.5 \times 10^{-15}$, depending on the length of the comparison.  For comparisons against the optical Hg$^+$ standard, statistical uncertainty can be made negligible ($\sim 2\times10^{-16}$) with less than 20~minutes of averaging, and the current calibration of Hg$^+$ to NIST-F1 is $0.65\times10^{-15}$.  The absolute frequency for all the comparisons is then derived by correcting the statistical value with the calibration of the reference against NIST-F1 and for the gravitational shift caused by relative height differences. 

\begin{figure}
\includegraphics[width=8.3cm]{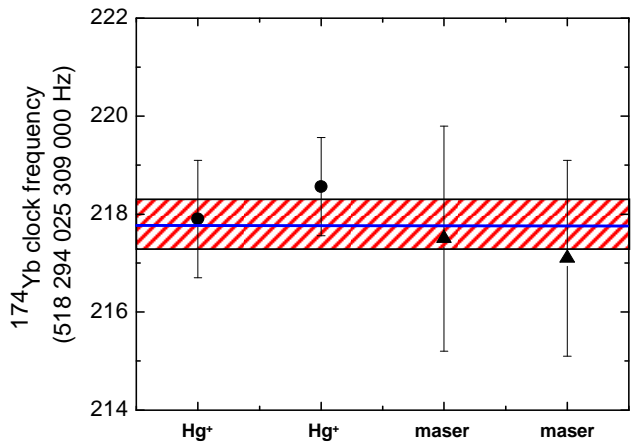}
\caption{Frequency evaluation of Yb clock transition.  Two measurements were against Hg$^+$ ($\bullet$) and the other two measurements were against a calibrated maser ($\blacktriangle$). The weighted mean of all the evaluations is $\nu(^{174}\textrm{Yb})= 518\,294\,025\,309\,217.8(0.9)$~Hz (shaded area is the one $\sigma$ confidence interval).}
\label{measfin}
\end{figure}

The final results of the frequency measurements of the clock transition in $^{174}$Yb are reported in Fig. \ref{measfin}.  Two values were obtained from optical comparisons against the Hg$^+$ standard and the other two were obtained though comparison against a NIST-F1 calibrated maser. The weighted mean of all the evaluations is $\nu$($^{174}$Yb) = 518\,294\,025\,309\,217.8(0.9)~Hz.

We have presented a frequency evaluation of the $^1S_0\leftrightarrow\!^3P_0$ clock transition in the $^{174}$Yb isotope at a level approaching that of the best atomic primary standards.  Fractional frequency uncertainty below that of the NIST-F1 standard ($\sim4\times10^{-16}$) will be acheivable in the near-term with further measurements of the systematic frequency shifts.  At this level, the current uncertainty in the room temperature BBR shift of Yb would become the limiting factor, and would need to be addressed for high accuracy ratio measurements against other optical standards~\cite{Rosenband08, Ludlow08}.  Then straightforward modifications to the experiment (i.e. better vacuum to improve the lifetime of the atoms in lattice, or multi-dimensional lattices) should lead to significant reductions in the other systematic frequency shifts and uncertainties.  Importantly, we foresee no significant barrier to achieving sub-$10^{-17}$ accuracy with an optical lattice clock based on magnetically induced spectroscopy of an even isotope of Yb.  

The authors would like to thank Jun Ye and the Sr clock team at JILA for our continuing collaboration.  A. Brusch acknowledges support from the Danish Natural Science Research Council.  L.S. Ma is supported by NSFC(60490280) \& STCSM(07JC14019).


\end{document}